\documentclass[letterpaper,english,aps,superscriptaddress,citeautoscript,reprint,nofootinbib,longbibliography]{revtex4-1}
\usepackage[T1]{fontenc}
\usepackage[latin9]{inputenc}
\usepackage{color}
\usepackage{babel}
\usepackage{units}
\usepackage{amsmath}
\usepackage{amssymb}
\usepackage{graphicx}
\PassOptionsToPackage{normalem}{ulem}
\usepackage{ulem}
\usepackage[unicode=true,pdfusetitle,
 bookmarks=true,bookmarksnumbered=false,bookmarksopen=false,
 breaklinks=false,pdfborder={0 0 0},pdfborderstyle={},backref=false,colorlinks=true]
 {hyperref}

\hypersetup{
 citecolor=blue,linkcolor=blue,urlcolor=blue}

\makeatletter

\pdfpageheight\paperheight
\pdfpagewidth\paperwidth

\pdfpageheight\paperheight
\pdfpagewidth\paperwidth

\newcommand{\be}{\begin{equation}}
\newcommand{\ee}{\end{equation}}
\newcommand{\bea}{\begin{eqnarray}}
\newcommand{\eea}{\end{eqnarray}}

\newcommand{\fig}[1]{Fig.~\ref{#1}}

\usepackage{fixltx2e}

\makeatother

\begin{document}
\title{Spectral properties of Co-decorated quasi 2-dimensional GaSe layer}
\author{I. Weymann}
\email{weymann@amu.edu.pl}

\affiliation{Faculty of Physics, Adam Mickiewicz University, 61-614 Pozna\'{n},
Poland}
\author{M. Zwierzycki}
\email{maciej.zwierzycki@ifmpan.poznan.pl}

\author{S. Krompiewski}
\affiliation{Institute of Molecular Physics, Polish Academy of Sciences, 60-179
Pozna\'{n}, Poland}

\date{\today}

\begin{abstract}
Based on reliable {\it ab initio} computations and the numerical renormalization group
method, systematic studies on a two-dimensional GaSe monolayer with
a Co adatom have been carried out. It is shown that the stable lowest-energy
configuration of the system involves the Co adatom located over Ga atom.
For such configuration, it is demonstrated that the electronic and magnetic
properties of the system can be effectively controlled by means of external factors,
such as magnetic field, gate voltage or temperature.
Moreover, if properly tuned, the GaSe-Co system can also exhibit the Kondo effect.
The development of the Kondo phenomenon is revealed in the local density of states of the Co adatom,
its magnetic field dependence, which presents the splitting of the Kondo peak,
as well as in the temperature dependence of the conductance,
which exhibits scaling typical of the spin one-half Kondo effect.
\end{abstract}

\maketitle

\section{Introduction}

Two-dimensional (2D) and quasi two-dimensional monolayers
have been recently intensively studied,
both experimentally and theoretically, because they exhibit many remarkable
physical phenomena, including mechanical, optoelectronic, electric
and magnetic ones \cite{Tiwari2016Jul,Li2016Jul,Novoselov2016Jul,Avouris2017Jul,Jappor2017Nov,Jappor2018Jul}.
Moreover, it is now also well known that such systems
can be additionally functionalized in various ways, e.g. by chemical
doping \cite{Cao:prl2015}, and by introducing either adatoms
or vacancies and other defects
\cite{Kaloni:prb14,Ao:PhysChemChemPhys2015,Lu:RSCAdv2017}.
In fact, the presence of magnetic adatoms can give rise
to interesting strongly correlated phenomena, such as the Kondo effect
\cite{Kondo_Prog.Theor.Phys32/1964,Hewson_book},
which still, despite a few decades passed
since its observation in artificial atoms \cite{Goldhaber_Nature391/98},
attracts a considerable attention \cite{Yoo2018Apr,Datta2019Sep,V.Borzenets2020Mar}.
Quite interestingly, the spectral signatures of adatoms placed 
on 2D materials have been analyzed in the Kondo regime
in a number of works \cite{Wehling2010Mar,Brar2010Oct,Ren2014Jul,
	Krychowski2014Jan,Weymann2017Sep}.
However, given a variety of 2D and quasi-2D materials,
there are sill some aspects that remain unexplored.

The system under consideration here
is the GaSe monolayer with a unit cell composed
of a Ga-Ga dimer covalently bonded to six Se atoms \cite{Cao:prl2015,Ma:PhChChPh2015}.
Similar to many other low dimensional graphene-like layers, this material
is also an object of particular interest. This is especially so because,
in contrast to graphene, GaSe is a semiconductor
with a pronounced energy gap---and can therefore serve as a transistor
and possibly also as a promising material for solar energy harvesting
\cite{Rawat:JPhysChemC2019}.
In this paper we in particular focus on examining the spectral properties
of GaSe monolayer decorated with Co adatoms.
The analysis is performed by combining the first-principle calculations 
with the non-perturbative numerical renormalization group (NRG) method 
\cite{Wilson_Rev.Mod.Phys.47/1975}.
The {\it ab initio} computations are used to determine the
density of states of the system as well as the electron occupation and 
magnetic moment of the Co adatom. Then, the spectral properties
of the adatom are determined by using NRG for an effective
Anderson-like Hamiltonian. 
It is shown that the electronic and magnetic properties
of the system can be tuned by external means,
such as temperature, magnetic field or gate voltage. 
Furthermore, it is also demonstrated that
the GaSe-Co system can exhibit the Kondo effect, if appropriately tuned.

\section{Electronic structure and effective model}
\subsection{First principles calculations}

The electronic structure of the system was determined using the full
potential linearized augmented plane wave method (FLAPW) \citep{singh:flapw2006}
as implemented in the WIEN2K package \citep{Blaha:wien2k}. In all
the calculations the generalized gradient approximation (GGA) of density
functional theory (DFT) was used with the exchange potential in the
Perdew-Burke-Ernzerhof parametrization \citep{Perdew_mz:prl96}.
The integration over 2D Brillouin zone was performed using
the mesh densities corresponding to several hundreds k-points (or
more) with the convergence criteria for energy, charge per atom and
forces set to $\unit[10^{-4}]{Ry}$, $\unit[10^{-3}]{e}$ and $\unit[1]{mRy/a.u.}$,
respectively. The GaSe monoloayers were separated by $\unit[14]{\mathring{A}}$
ensuring the lack of hopping between the neighboring planes.

\begin{figure}[th]
\begin{centering}
\includegraphics[width=0.9\columnwidth]{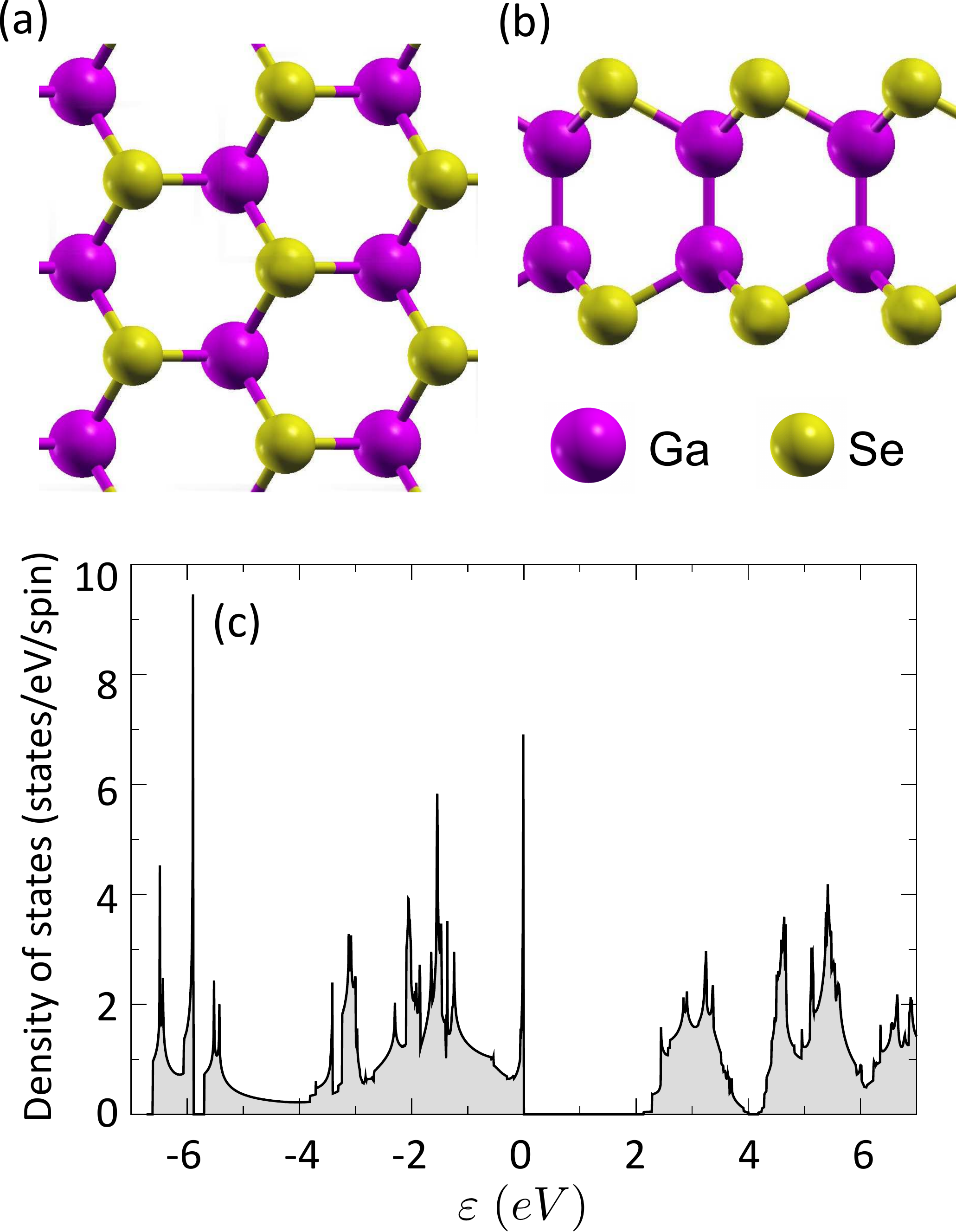} 
\par\end{centering}
\caption{The structure of the GaSe monolayer as seen (a) from the top and (b)
from the side. (c) The density of states of unadorned GaSe monolayer.
\label{fig:GaSe-structure}}
\end{figure}

The structure of GaSe monolayer, consisting of two atomic layers,
is shown in Figs.\,\ref{fig:GaSe-structure}(a)-(b). In the first
step the lattice constant and the internal positions of 4 atoms within
the unit cells were determined, the resulting parameters being $a=3.755,$
$d_{Ga-Ga}=2.$46 and $d_{Se-Se}=\unit[4.85]{\text{Å}}$ for the lattice
constant and the distances between Ga and Se atoms, respectively.
The calculated values are in good agreement with both the experimental
data \citep{Li:SciRep2014} and previous calculations \citep{Ma:PhChChPh2015}.
The density of states (DOS) calculated for isolated monolayer is shown
in Fig.~\ref{fig:GaSe-structure}(c). The value of clearly observable
band gap of $\Delta=\unit[2.1]{eV}$ is close to previously calculated
values \citep{Ma:PhChChPh2015,Cao:prl2015}.

\begin{figure}
\begin{centering}
\includegraphics[width=0.9\columnwidth]{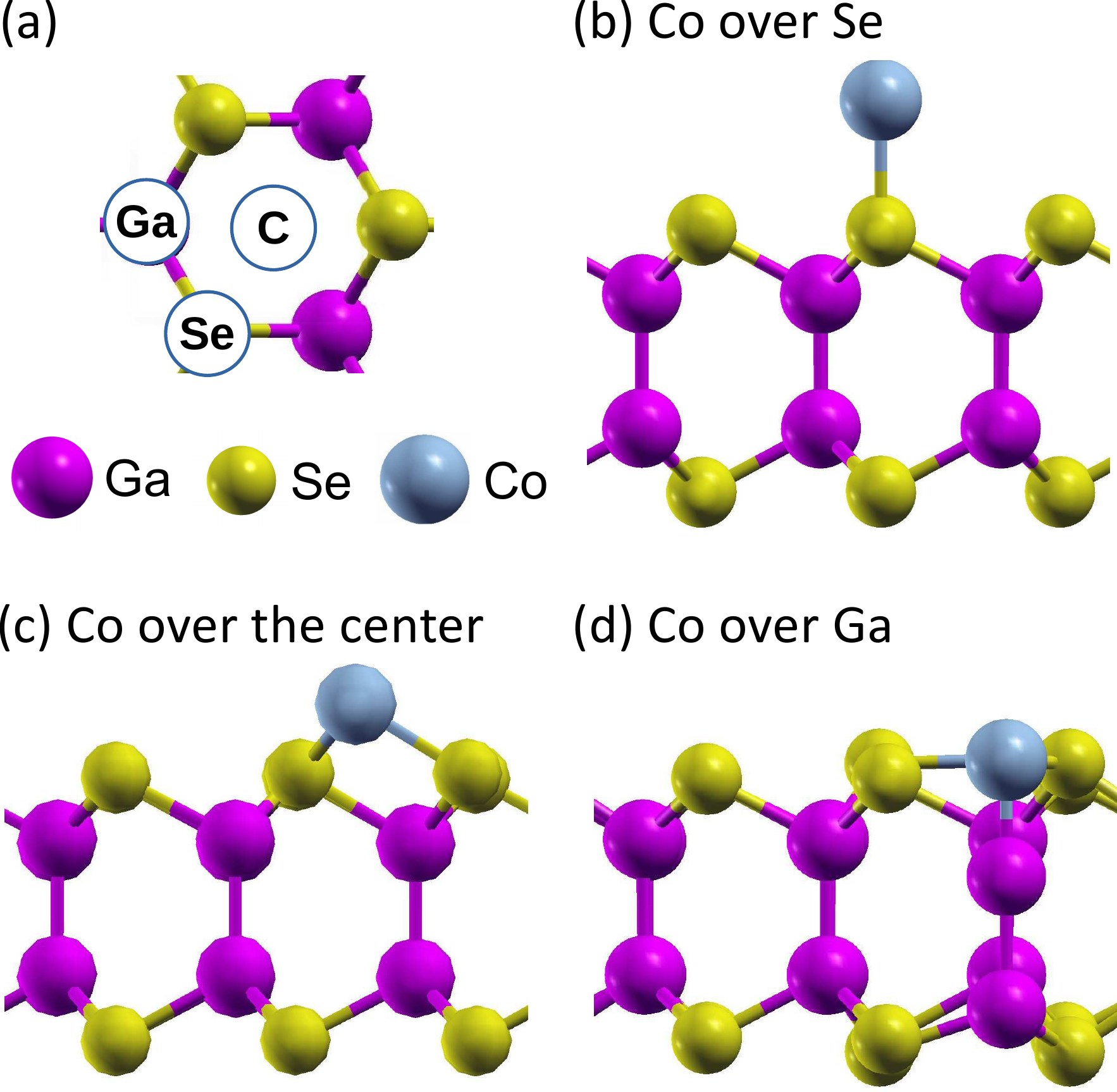} 
\par\end{centering}
\caption{(a) Three possible positions of Co adatom -- over Ga, Se and over
the center of the hexagon (marked as C) -- as seen from the top.
(b)-(d) Side views of the cases from (a) (relaxed structures).\label{fig:GaSeCo-structure}}
\end{figure}

The case of GaSe monolayer with Co adatom was studied using 3x3 supercell
containing 37 atoms (18 Ga, 18 Se and one Co atom). Within the supercell,
the size of which was fixed to the multiple of the previously determined
values, the internal positions were fully relaxed. The place of most
probable attachment of the adatom was determined by comparing the
total energies for three high symmetry cases with Co positioned over
Se, Ga and the center of the hexagon. These are indicated in Fig.
\ref{fig:GaSeCo-structure}(a), while the side views of the relaxed
structures are shown in the panels (b)-(d) of the same figure. Visual
analysis of the three structures suggests that the bonding strength
is the weakest in the case of Co over Se {[}Fig. \ref{fig:GaSeCo-structure}(b){]}
and the strongest for Co over Ga {[}Fig. \ref{fig:GaSeCo-structure}(d){]}.
The Se-Co distance in the first case equals $\unit[2.46]{\text{Å}}.$
In the second case of centrally located adatom, its height over the
plane of Se atoms is visibly smaller ($\unit[1.27]{\text{Å}}$).
On the other hand, in the third case, Co over Ga,
the adatom is actually embedded in the Se plane.

This conclusion is supported by a direct comparison of the total energy
values, which stand in the following order 
\[
E_{Ga}<E_{center}<E_{Se},
\]
with the lowest configuration being well separated from the other
two ($E_{center}-E_{Ga}\approx\unit[6]{eV}$). In comparison, the
difference between the second and the third configuration energy is
an order of magnitude smaller ($E_{Se}-E_{center}\approx\unit[0.7]{eV}$).
The total magnetic moment, concentrated on the Co atom, ranges between
atomic-like $\unit[3]{\mu_{B}}$ for the weakly bonded ``Co over
Se'' geometry to the substantially reduced $\unit[1]{\mu_{B}}$ in
the lowest energy case.

\begin{figure}[t]
\begin{centering}
\includegraphics[width=0.95\columnwidth]{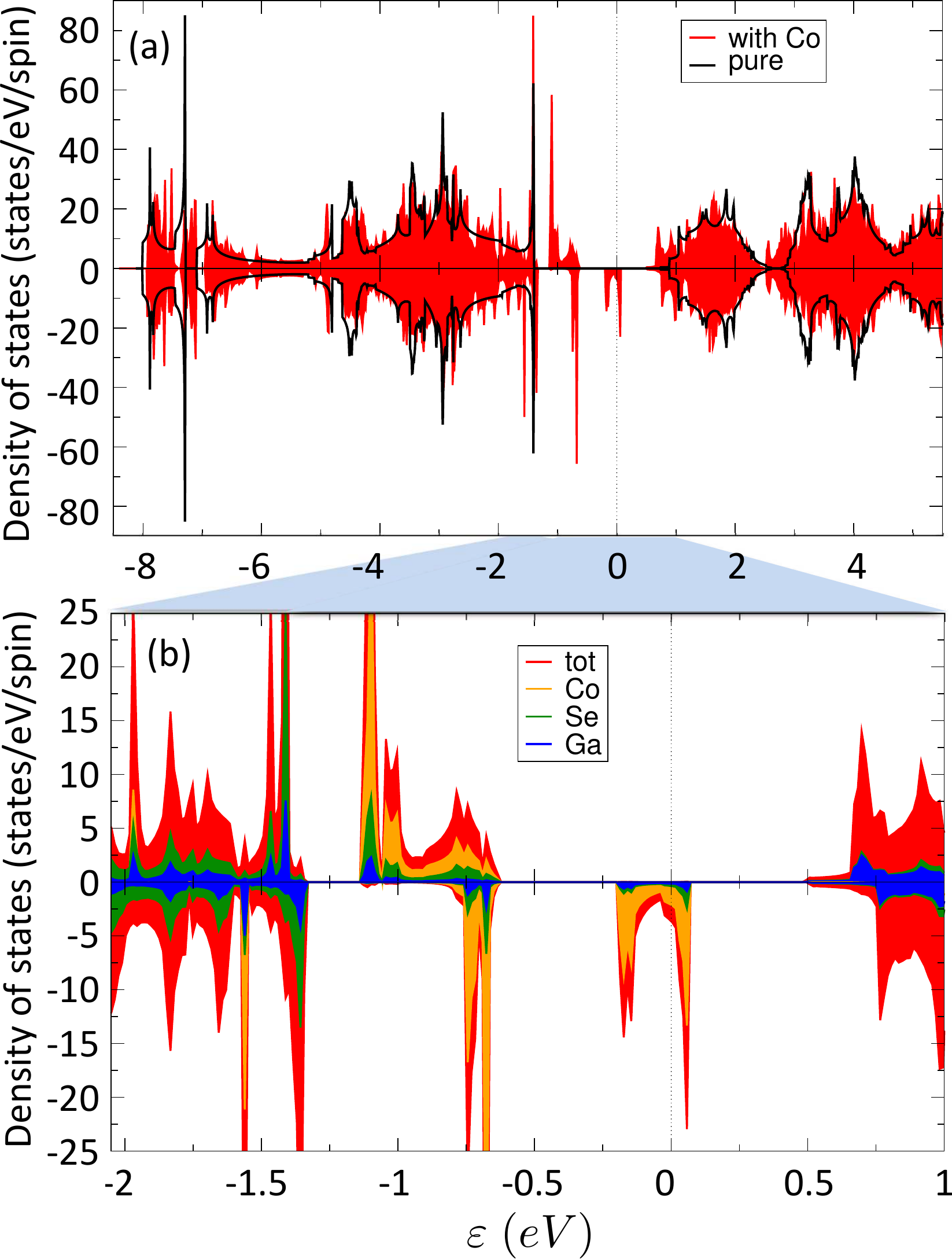} 
\par\end{centering}
\caption{ (a) The density of states for GaSe with Co adatom (red) together
with the density of states of clean GaSe (black line). (b) The closeup
of the behavior of the density of states around the Fermi energy,
where the atomic contributions are also presented. \label{fig:DOS_with_Co}}
\end{figure}

The density of states for the system with Co adatom, shown in Fig.~\ref{fig:DOS_with_Co}(a)
using red color, follows for the most part an outline of the density
of states for clean GaSe (black line). The most visible differences
between the two densities are located in the energy window corresponding
to the clean case gap.\footnote{One notes that the Fermi energy of the Co-decorated
system is anchored within the region of original gap because of the
presence of the localized states discussed in the text.}
This is shown in Fig. \ref{fig:DOS_with_Co}(b) where in addition
to the total DOS (red), the atomic contributions are also shown. The
latter are, in the case of Se (green) and Ga (blue), summed over all
the atoms of the given kind within the supercell. The highly localized
states visible within the region of the GaSe gap are of almost pure
Co$-d$ character (predominantly $d_{x^{2}+y^{2}},d_{xy},d_{xz}$
and $d_{yz}$), although certain level of hybridization with the surrounding
atoms of the host can also be deduced. 

In conclusion, the $\unit[1]{\mu_{B}}$ value of the magnetic moment
coupled with its almost complete localization on adatom, suggests
that the low-energy properties of the system under consideration can
be correctly described using an effective spin-$\frac{1}{2}$ Anderson impurity
model \citep{Anderson1961Oct}, as discussed in the following subsection.

\subsection{Effective model}

To obtain the most accurate results for the spectral properties of
the Co adatom, we employ the density-matrix numerical renormalization
group method \citep{Wilson_Rev.Mod.Phys.47/1975,Legeza_DMNRGmanual},
developed originally by Wilson for the Kondo problem \citep{Hewson_book}.
The effective Anderson-like model for the NRG computations has been
constructed following the method described in Refs.\,\citep{Ujsaghy2000Sep,Weymann2017Sep}.
The crucial parameters of the Co adatom in this context are: the on-site
Coulomb repulsion (U), the level energy ($\epsilon$), the hybridization
parameter ($V_{pd\sigma}$), and the Co $d$-shell occupancy ($n_{d}$).
The latter has been directly found from the \textit{ab initio} calculations,
to be equal to $n_{d}=7.6$. The parameter $U$ for Co is equal to
$\unit[4]{eV}$ \citep{Kaloni:prb14}, whereas the hybridization has been found
to be $V_{pd\sigma}=-0.83$eV by using the Harrison's scaling method
\citep{Harrison:electronic1980} for the computed distances $R_{CoGa}=2.16$Å
(1st nearest neighbor) and $R_{CoSe}=2.33$Å (2nd nearest neighbors).
The charge states of the cobalt $d$-level correspond
to energies \citep{Ujsaghy2000Sep} 
\begin{equation}
E(j)=j\epsilon+Uj(j-1)/2\label{eq:d-energy}
\end{equation}
with the minimum value for $j_{min}\equiv n_{d}=1/2-\epsilon/U$,
\textit{i.e.} $\epsilon=-(n_{d}-1/2)U.$

The energy separations of the relevant levels $j=7,8,9$ are $\tilde{\epsilon}_{d}=E(8)-E(7)$
and $2\tilde{\epsilon}_{d}+\tilde{U}=E(9)-E(7)$ and determine the
parameters of the effective Anderson-like Hamiltonian. One then finds,
$\tilde{U}=U$, and $\tilde{\epsilon}_{d}=\epsilon+7U=U(7.5-n_{d})$.
In calculations, we perform logarithmic discretization of the density
of states of pure GaSe. Then, we tridiagonalize the discretized Hamiltonian
by Lanczos method to obtain the hoppings and on-site energies along
the Wilson chain \citep{Wilson_Rev.Mod.Phys.47/1975}. In the next
step, the eigenenergies and eigenvectors of the Hamiltonian are determined
in an iterative fashion, which allow for determination of the density
matrix and the relevant spectral operators. To perform computations,
we have assumed the discretization parameter equal to $\Lambda=1.8$
and kept at least $1500$ states at each iteration step. The quantity
of interest is the spectral function $A(\omega)$ of the adatom, which
is defined as $A(\omega)=-{\rm Im}\left[G^{R}(\omega)\right]/\pi$,
where $G^{R}(\omega)$ is the Fourier transform of the retarded Green's
function of the adatom's effective orbital level.

\section{Transport properties}

\begin{figure}[t]
\begin{centering}
\includegraphics[width=1\columnwidth]{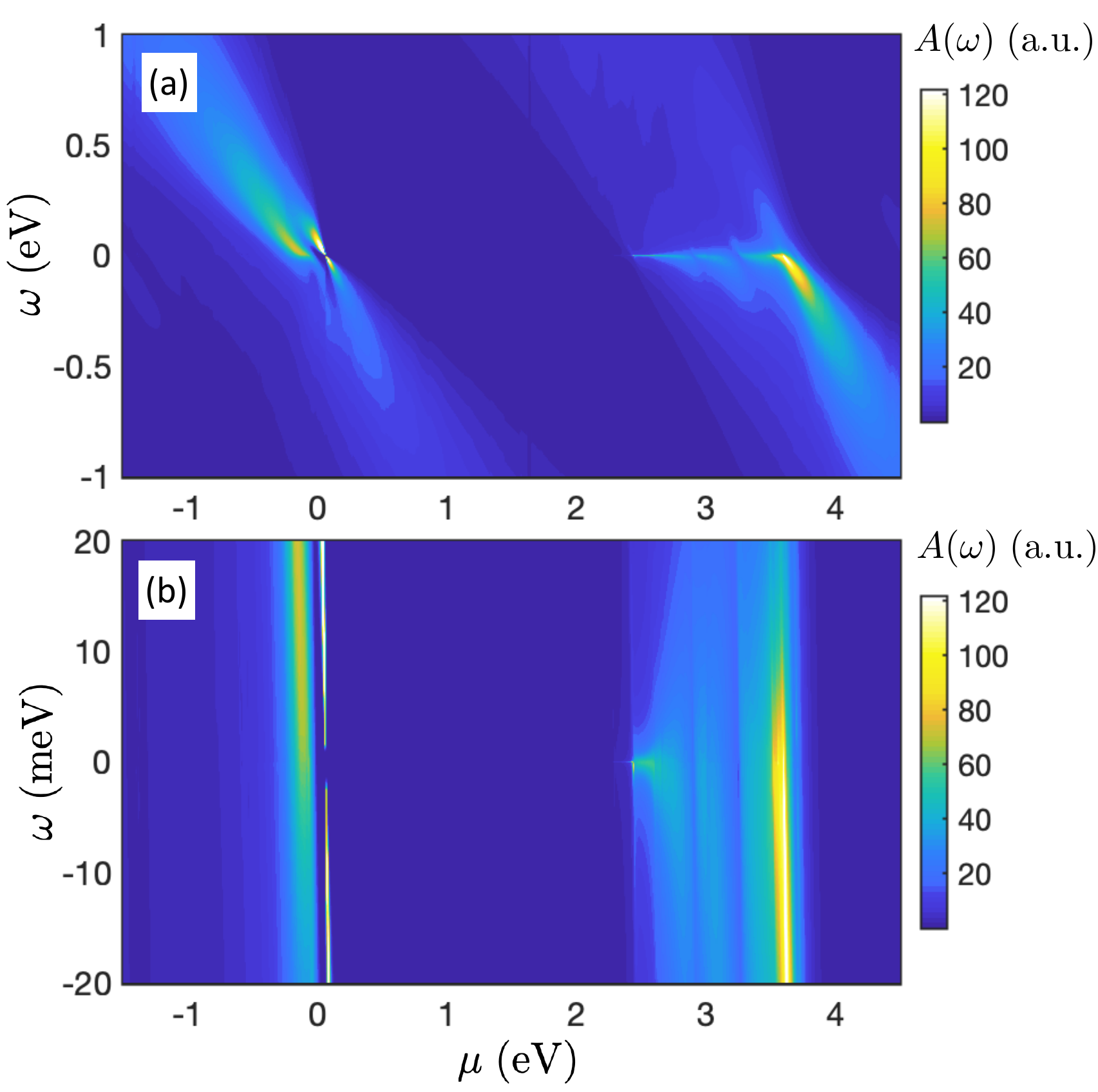} 
\par\end{centering}
\caption{ (a) The zero-temperature spectral function of the Co adatom calculated
as a function of chemical potential $\mu$. (b) The low-energy behavior
of the spectral function. The parameters of the effective Hamiltonian
are $\tilde{U}=4$ eV, $\tilde{\epsilon}_{d}=-0.4$ eV and hybridization
$V_{pd\sigma}=-0.83$ eV. \label{fig:A2D}}
\end{figure}

\begin{figure}[t]
\begin{centering}
\includegraphics[width=0.9\columnwidth]{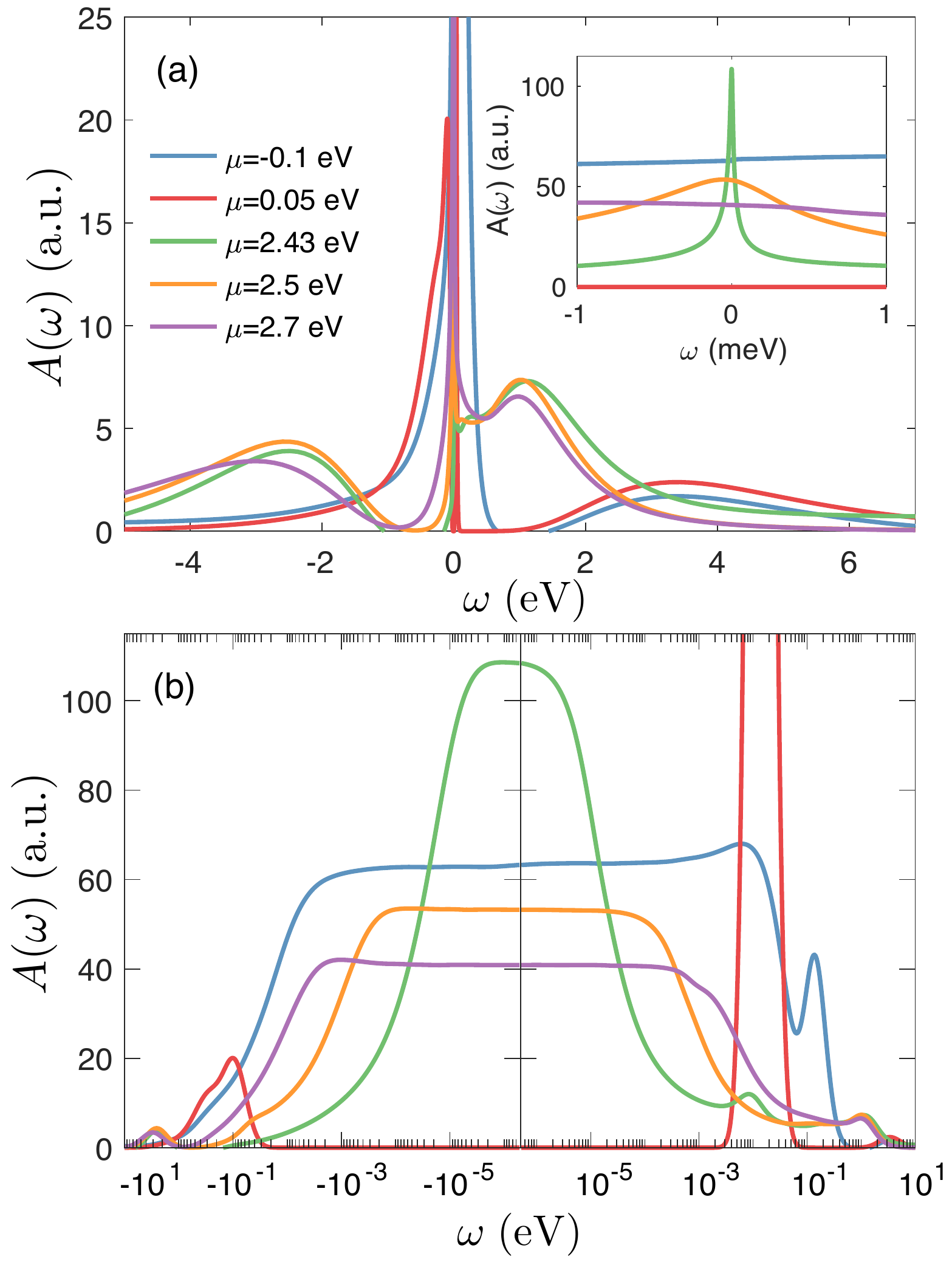} 
\par\end{centering}
\caption{ The spectral function for selected values of the chemical potential
$\mu$ plotted in the (a) linear and (b) logarithmic scale. The inset
in (a) presents the low-energy behavior of the spectral function where
a considerable Kondo peak is visible. The parameters are the same
as in Fig.~\ref{fig:A2D}. \label{fig:A1D}}
\end{figure}

The spectral function of Co adatom $A(\omega)$ as a function of the
chemical potential $\mu$ and energy $\omega$ is shown in \fig{fig:A2D}(a),
while the low-energy behavior of $A(\omega)$ is displayed in \fig{fig:A2D}(b).
Because the density of states of pure GaSe exhibits a gap of the order
of $2.1$ eV, see \fig{fig:GaSe-structure}(c), the spectrum has
negligible weight at low-energies in the corresponding regime of $0.2{\rm eV}\lesssim\mu\lesssim2.3{\rm eV}$.
When the chemical potential is detuned out of this regime, the spectral
function exhibits low-energy features resulting from strong correlations
between the band electrons and the spin localized in the adatom's
orbital. One of such phenomena is the Kondo effect, which manifests
itself through a zero-energy resonance in the local density of states
\citep{Kondo_Prog.Theor.Phys32/1964,Hewson_book}. Indications of
such resonances can be clearly seen in Figs. \ref{fig:A2D}(a)-(b).

Moreover, the low-energy behavior of Co adatom is also revealed in
\fig{fig:A1D}, which shows the energy-dependence of $A(\omega)$
plotted both on the linear and logarithmic scale for selected
values of the chemical potential. One can clearly identify the Hubbard
resonances corresponding to $\omega+\mu\approx\tilde{\epsilon}_{d}$
and $\omega+\mu\approx2\tilde{\epsilon}_{d}+\tilde{U}$. Interestingly,
for all considered values of $\mu$, except for $\mu=0.05$ eV, there
is a zero-energy resonance in the spectral function. This peak is
clearly visible in the close-up on the low-energy behavior of $A(\omega)$
shown in \fig{fig:A1D}(a) as well as in the logarithmic-scale dependence
presented in \fig{fig:A1D}(b). Because for the selected values of
chemical potential the adatom is mostly occupied by a single electron,
one can conclude that the resonance is due to the Kondo effect. Here,
however, one needs some care, since not all resonances need to be
due to the Kondo correlations.

\begin{figure}[t]
\begin{centering}
\includegraphics[width=0.9\columnwidth]{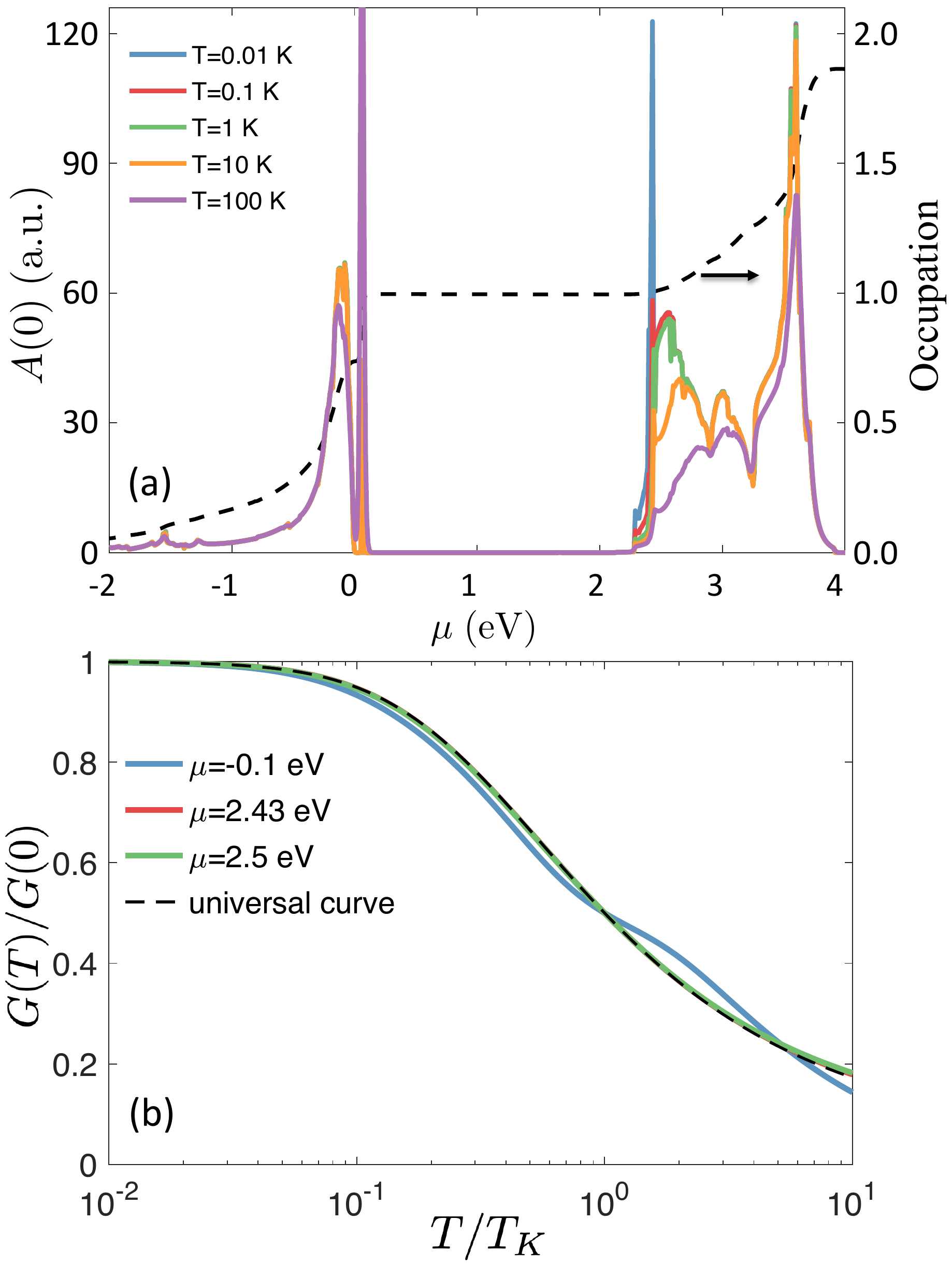} 
\par\end{centering}
\caption{ (a) The spectral function at the Fermi energy $A(0)$ calculated
for several values of temperature, as indicated. The dotted line presents
the occupation of the adatom's orbital level. (b) The temperature
dependence of the normalized linear conductance $G(T)/G(0)$ calculated
for selected values of the chemical potential as a function of the
normalized temperature $T/T_{K}$. The dashed line presents the universal
scaling plot for the spin-1/2 Kondo effect. The parameters are the
same as in Fig.~\ref{fig:A2D}. \label{fig:A0}}
\end{figure}

To shed more light on this issue, in \fig{fig:A0}(a) we present
the chemical potential dependence of the spectral function taken at
the Fermi energy $A(0)$. Because the Kondo resonance develops for
temperatures $T$ lower than the Kondo temperature $T_{K}$, $T\lesssim T_{K}$,
a strong dependence of $A(0)$ on $T$ can be observed for $\mu\lesssim0.2$
eV and $\mu\gtrsim2.3$eV. Nevertheless, as mentioned before, a special
caution is needed since some low-energy features can be related to
enhanced charge fluctuations around $\mu\approx\tilde{\epsilon}_{d}$ and
$\mu\approx \tilde{\epsilon}_{d}+\tilde{U}$. More specifically, the effects
associated with Kondo correlations develop only in the local moment
regime \cite{Hewson_book}, i.e. $\tilde{\epsilon}_{d}\lesssim\mu\lesssim\tilde{\epsilon}_{d}+\tilde{U}$,
where the adatom's occupation is odd, see the dotted line in \fig{fig:A0}(a).
To understand the origin of the low-energy resonances, we have determined
the linear response conductance at different temperatures, which would
correspond to e.g. measuring the conductance over the Co adatom by STM.
The normalized conductance in linear response can be found from \cite{MeirWingreen}
$G(T)/G(0) = \int d\omega [-f'(w)]A(\omega)/A(0)$,
where $f(\omega)$ denotes the Fermi-Dirac distribution function.
In \fig{fig:A0}(b) we display the temperature dependence of
$G(T)/G(0)$ as a function of the normalized temperature $T/T_{K}$,
where $T_{K}$ is the Kondo temperature
defined as temperature at which $G(T)=G(0)/2$.
One can see that the resonances for $\mu=2.43$eV and $\mu=2.5$eV have the scaling characteristic
of the Kondo effect---$G(T)/G(0)$ follows then exactly the universal
scaling dependence of the spin-1/2 Kondo effect \cite{Hewson_book}.
However, $G(T)/G(0)$ in the case of $\mu=-0.1$eV does not match with the Kondo universal
scaling and one can conclude that this resonance is not due to Kondo
correlations, but rather associated with resonant charge fluctuations.

\begin{figure}[t]
\begin{centering}
\includegraphics[width=1\columnwidth]{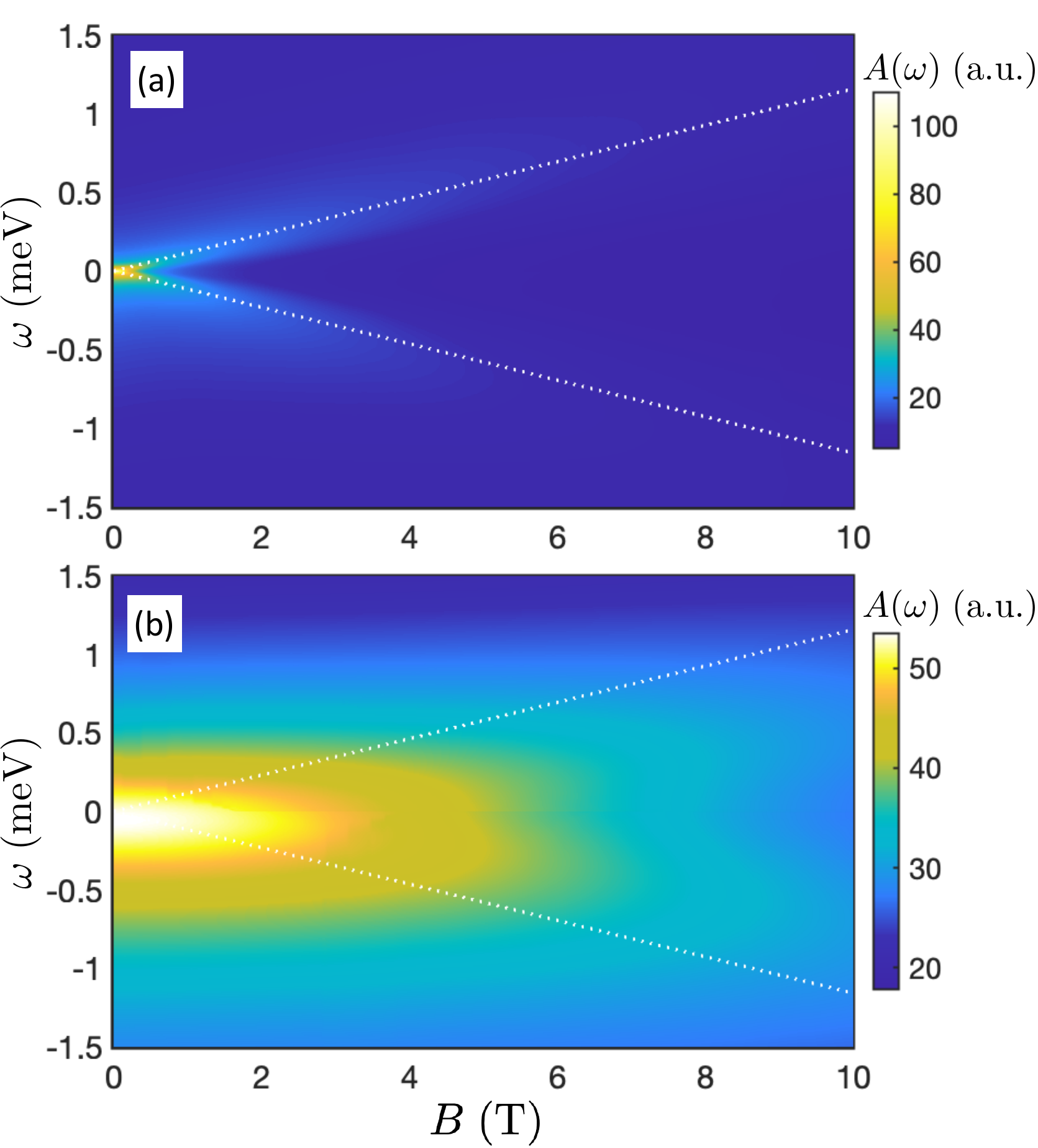} 
\par\end{centering}
\caption{ Spectral function plotted vs magnetic field demonstrating the splitting
of the Kondo resonance for (a) $\mu=2.43$ eV and (b) $\mu=2.5$ eV.
The parameters are the same as in Fig.~\ref{fig:A2D} and the g-factor
was assumed to be $g=2$. The dotted lines indicate the Zeeman splitting
energy $\Delta_{Z} = g\mu_B B$. \label{fig:B}}
\end{figure}

To corroborate our observations even more, in \fig{fig:B} we have
determined the magnetic field dependence of the spectral function
for two selected values of chemical potential, for which the temperature
scaling of the conductance revealed the Kondo origin.
Indeed, it can be clearly seen that the Kondo resonance becomes split by magnetic
field and only side resonances at energies corresponding to the Zeeman
energy $\omega\approx\pm\Delta_{Z}$ are present, where
$\Delta_Z = g\mu_B B$, see the dotted lines in \fig{fig:B}.
Note also that in the case of $\mu=2.5$ eV a larger magnetic field is necessary to
suppress and split the Kondo resonance, which is associated with larger
Kondo temperature in this case compared to $\mu=2.43$ eV.

\section{Conclusions}

In this paper we have investigated electronic and magnetic properties
of the GaSe monolayer with a cobalt adatom. The focus has been put
on the detailed analysis of local densities of states, spectral functions,
and the possible Kondo resonances and their evolution with external magnetic
field. The studies have been carried out in three steps by combining:
(i) the first principles calculations, (ii) construction of an effective
Anderson-type Hamiltonian, and (iii) the application of the numerical
renormalization group. This allowed us for obtaining very accurate predictions
for the spectral properties of Co-decorated quasi 2-dimensional GaSe layer.

In particular, the first-principle calculations provided the information about the 
lowest-energy configuration, the magnetic moment and occupation of the Co adatom.
This information was further exploited to determine the parameters
of the effective Hamiltonian and perform the NRG calculations.
It turns out that the calculated local density of states
of the Co adatom exhibits resonances depending on the position of the chemical potential.
We have shown that these resonances can be associated with the Kondo
correlations that develop between the adatom and two-dimensional host material.
This observation has been corroborated by
the analysis of the system's transport properties in
external magnetic field and finite temperatures.
In particular, the splitting of the Kondo peak in magnetic field 
as well as the Kondo universal scaling of the linear conductance have been demonstrated.

We believe that our study sheds more light on the 
spectral properties of magnetic adatom-decorated two dimensional materials
and, especially, on the Kondo phenomena in such systems.
The results show that the Co-decorated GaSe monolayer
is an attractive Kondo system and its properties can be 
tuned by gate voltage or external magnetic field.


\begin{acknowledgments}
	I.W. acknowledges support by the Polish National Science
	Centre from funds awarded through the decision No. 2017/27/B/ST3/00621.
	The computing time at the Pozna\'n Supercomputing
	and Networking Center is also acknowledged.
\end{acknowledgments}



%

\end{document}